\documentclass[reprint,superscriptaddress]{revtex4-2}

\usepackage[colorlinks=true, linkcolor=blue, citecolor=blue, urlcolor=blue]{hyperref}

\usepackage{graphicx}
\usepackage{bm}% bold math
\usepackage{bbm}
\usepackage{bbold}
\usepackage{amsmath}
\usepackage{braket}
\usepackage{color}
\usepackage[normalem]{ulem}
\usepackage[caption=false]{subfig} % REVTeX-compatible
\usepackage{pdfrender}  % more text control
\usepackage[percent]{overpic}  % text over figures
\usepackage{ifthen}

\newcommand{\cut}[1]{} % Use this line to hide deleted text

\begin{document}

\title{Experimental Realization of Rabi-Driven Reset for Fast Cooling of a High-Q Cavity}

\author{Eliya Blumenthal}
\thanks{Corresponding author. email: eliyab@campus.technion.ac.il}
\author{Natan Karaev}
\author{Shay Hacohen-Gourgy}
\affiliation{Department of Physics, Technion - Israel Institute of Technology, Haifa 32000, Israel}

\begin{abstract}
    High–$Q$ bosonic memories are central to hardware–efficient quantum error correction, but their very isolation makes fast, high–fidelity reset a persistent bottleneck. Existing approaches either rely on weak intermode cross–Kerr conversion or on measurement–based sequences with substantial latency. Here we demonstrate a hardware–efficient \emph{Rabi–Driven Reset} (RDR) that implements continuous, measurement–free cooling of a superconducting cavity mode. A strong resonant Rabi drive on a transmon, together with sideband drives on the memory and readout modes detuned by the Rabi frequency, converts the dispersive interaction into an effective Jaynes–Cummings coupling between the qubit’s dressed states and each mode. This realizes a tunable dissipation channel from the memory to the cold readout bath. Crucially, the engineered coupling scales with the dispersive interaction to the qubit and the drive amplitude, rather than with the intermode cross–Kerr, enabling fast cooling even in very weakly coupled architectures that deliberately suppress direct mode–mode coupling. We demonstrate RDR of a single photon with decay time of $1.2\,\mathrm{\mu s}$, more than two orders of magnitude faster than the intrinsic lifetime.  Furthermore, we reset $\sim30$ thermal photons in $\sim 80 \mathrm{\mu s}$ to a steady-state average photon number of $\bar{n}= 0.045\pm0.025$.
\end{abstract}

\maketitle

\section*{Introduction}
Continuous-variable bosonic modes provide a promising platform for encoding quantum information and realizing error-correctable quantum computers. Recent progress has demonstrated several bosonic codes, such as the cat code\,\cite{DissipativeCat} and the GKP code\,\cite{ECDCooling}. Preparing and manipulating bosonic codes requires extensive calibration and repeated measurements. At the same time, Modes with long coherence times demand strong isolation from the environment\,\cite{highQCavity}, which introduces an inherent tradeoff: one must often wait for the system to naturally relax to its steady state before performing the next operation. In contrast with two level systems, projective measurements can not solve this problem for bosonic-modes due to their large Hilbert spaces. Actively accelerating dissipation, that is, cooling the system more rapidly, provides a practical solution to this challenge.

We will focus on active cooling of bosonic-modes in circuit quantum electrodynamics systems. In these architectures, the high-Q bosonic-mode (memory mode) is typically implemented as either a planar resonator or a three-dimensional cavity. The memory mode is dispersively and weakly coupled to a transmon qubit, enabling universal control\,\cite{ECD}, while the qubit itself is coupled to a dissipative readout resonator for measurement.

Active cooling of memory modes has been realized using three main approaches. The first two involves coupling the memory mode and the readout mode by modulating the cosine potential of the Josephson-Junction. This was done by driving both modes slightly off resonance\,\cite{BadCooling} or by driving the transmon at the mean of the modes frequencies\,\cite{highQCavity}. The cooling rate of the first method scales as the cross-Kerr between the modes, making it very slow for weakly coupled systems, and the second method is hard limited in drive power to prevent higher excitations of the transmon. The third approach employs repeated cycles in which excitations from the memory mode are swapped into the qubit using conditional displacements and qubit rotations, followed by resetting the qubit\,\cite{ECDCooling}. While effective, this scheme requires high-fidelity readout, and each reset cycle is relatively long since it consists of two conditional displacement gates and a complete qubit measurement. This method was also followed by repeated projective measurements onto the vacuum state to achieve high fidelity. Such projections must be employed after significant cooling of the memory mode, and can also be integrated with the other methods.
\begin{figure}
    \centering
    \begin{overpic}[width=0.99\linewidth]{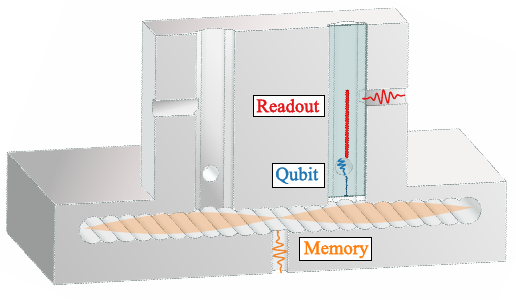}
    \put(-1,55){(a)}
    \end{overpic}
    \begin{overpic}[width=0.45\linewidth]{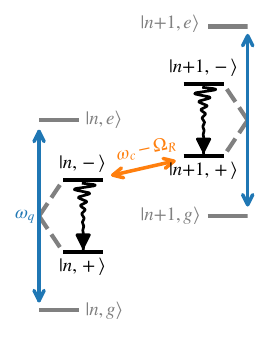}
    \put(-1,90){(b)}
    \end{overpic}
    \hspace{5mm}
    \begin{overpic}[width=0.45\linewidth]{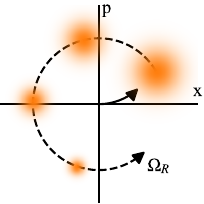}
    \put(-1,111){(c)}
    \end{overpic}
    
    \caption{Experiment scheme. (a) A cross-sectional view of a high-Q flute cavity\,\cite{fluteCavity2021}, where the TE102 mode is dispersively coupled to a transmon qubit, which in turn is coupled to a stripline readout resonator. The sideband drives for each mode and the Rabi drive are applied through designated ports (For the full wiring diagram see Appendix\,\ref{appendix:system}.). The cavity includes room for an additional qubit, left vacant in this experiment. (b) Illustration of the sideband cooling mechanism. The orange arrow indicates sideband coupling between the memory mode and the Rabi-driven qubit’s dressed states. The undulating line denotes qubit decay through the readout mode. (c) Wigner functions of a thermal state actively cooling down in frame displaced by $-\bar{a}_me^{\mathrm{i}\Omega_Rt}$.}
    \label{fig:Scheme}
\end{figure}

In this work, we demonstrate a fast memory mode sideband-cooling technique, first proposed in Ref.\,\cite{NatanTheory}, that extends a cooling protocol originally developed for qubits\,\cite{MurchBath}. Specifically, we apply a strong Rabi drive to the qubit together with sideband drives on both the memory- and readout modes, detuned by the Rabi frequency $\Omega_R$. Starting from a dispersive coupling
\begin{equation}
    H_\mathrm{Dispersive}=\sum_{i=\{m,r\}} \chi_i a_i^\dagger a_i\sigma_z,
\end{equation}
where subscript m marks the memory mode and subscript r marks the readout mode, $2\chi_i$ is the dispersive shift and $a_\mathrm{i}$ is the annihilation operator of mode i, and $\sigma_z$ is the Pauli-Z operator of the qubit, these drives generate an effective Jaynes–Cummings (JC) interaction between the qubit’s dressed states and each mode in a displaced frame (depicted in Fig.\,\ref{fig:Scheme}(b)). The displaced frame is described by the unitary transformation $U(t) = D_m(-\bar{a}_me^{\mathrm{i}\Omega_Rt})D_r(-\bar{a}_re^{\mathrm{i}\Omega_Rt})$, where $\bar{a}_i$ is the amplitude of the coherent state induced by the sideband drive and $D_i(\alpha)=e^{\alpha a_i^\dagger-\alpha^*a_i}$ is the displacement operator of mode i. The effective Hamiltonian is given by
\begin{equation}
    H_\mathrm{effective} = \sum_{i=\{m,r\}} \chi_i \bar{a}_i \left(\sigma_+ a_i + \sigma_- a_i^\dagger\right),
\end{equation}
where $\sigma_-$ is the lowering operator of the qubit. This effective Hamiltonian is valid in the regime where the Rabi frequency is large compared with the dispersive shift $\Omega_R\gg\chi_\mathrm{i},\chi_\mathrm{i}\bar{a}_\mathrm{i}$, but small compared to the anharmonicity of the transmon. As a result, the memory mode becomes coupled to the readout mode (and its dissipative environment) through the qubit, enabling efficient energy dissipation. In particular, the memory mode decays to a rotating coherent state (depicted in Fig.\,\ref{fig:Scheme}(c)), which is then displaced to the vacuum state when the sideband drive is ramped-down.

Our approach, which we call the \emph{Rabi-Driven Reset} (RDR), demonstrates cooling that is faster than previous methods as it generates continuous dissipation without compromising high cooling rate. The key is that the coupling rate scales with the dispersive coupling rate between each mode and the qubit, and not as the cross-Kerr coupling rate between the modes, which is designed to be negligible.

\section*{Results}
We implemented RDR on an electromagnetic mode (see Fig.\,\ref{fig:Scheme}(a)) with frequency $\omega_m/2\pi = 6.914\,\mathrm{GHz}$, which has a single-photon lifetime of $1/\kappa_m = 170\,\mathrm{\mu s}$. This mode is dispersively coupled to a transmon qubit with strength $2\chi_m/2\pi = 57\,\mathrm{kHz}$. The first transition frequency of the transmon is $\omega_q/2\pi = 6.33\,\mathrm{GHz}$ and its anharmonicity is $E_C/2\pi = 265\,\mathrm{MHz}$. The qubit relaxation time is $T_1=25\,\mathrm{\mu s}$ and echoed dephasing time is $T_2^\mathrm{Echo}=20\,\mathrm{\mu s}$. The readout mode has a resonance frequency of $\omega_r/2\pi = 7.7\,\mathrm{GHz}$, a linewidth of $\kappa_r/2\pi = 0.382\,\mathrm{MHz}$, and a dispersive shift of $2\chi_r/2\pi = 0.635\,\mathrm{MHz}$.

We calibrated the drives to apply RDR as follows: We chose $\Omega_R/2\pi = 9\,\mathrm{MHz}$, which exceeds all other parameters in the system.  We then calibrated the drive amplitudes, $\varepsilon_{m}/2\pi$ and $\varepsilon_r/2\pi$, assuming
\begin{equation}\label{eq:abar}
    \bar{a}_\mathrm{i} = \frac{\varepsilon_\mathrm{i}}{\mathrm{i}\Omega_R -\kappa_\mathrm{i}/2}\approx \frac{\varepsilon_\mathrm{i}}{\mathrm{i}\Omega_R}.
\end{equation}
This was done by measuring the drive-induced Stark-shift in a Ramsey experiment while the drive was applied (see Appendix\,\ref{appendix:calibration}). We then measured the Stark-shift induced by both drives applied in parallel. Lastly, we calibrated $\Omega_R$ in the doubly Stark-shifted frame by performing a Rabi experiment (see Appendix\,\ref{appendix:Rabi}).
\begin{figure}
    \centering
    \subfloat{\begin{overpic}[width=0.99\linewidth]{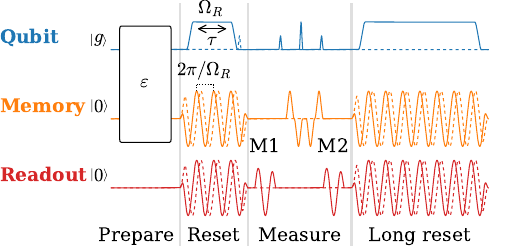}
    \put(-1,47.5){(a)}
    \end{overpic}}\\[5mm]
    \subfloat{\begin{overpic}[width=0.45\linewidth]{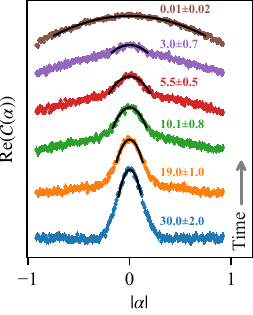}
    \put(0,105){(b)}
    \end{overpic}}
    \hspace{5mm}
    \subfloat{\begin{overpic}[width=0.45\linewidth]{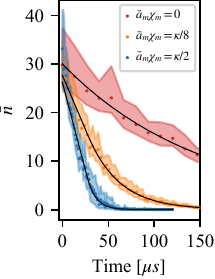}
    \put(0,105){(c)}
    \end{overpic}}
    
    \caption{Reset of a thermal state. (a) Pulse sequence for reset and measurement of arbitrary coherent or incoherent states. Solid (dashed) lines correspond to the real (imaginary) parts of the pulses. The $M1$ measurement (applied using an intrinsic readout mode reset pulse shape\,\cite{drachma}) is used to post-select the state of the qubit, as in Ref.\,\cite{ECD} and $M2$ is the actual characteristic function measurement. (b) Wigner characteristic function averaged over the real and imaginary axes of $\alpha$. Different times are vertically shifted for clarity. The fit considered only points above a threshold between 0.7 and 0.85. (c) Average photon number as a function of time for active reset compared to the natural decay of the memory mode. The shaded regions signify uncertainty about $\bar{n}$ due to fitting of different number of data points with respect to different thresholds. The active reset data is fitted to a piecewise linear-exponential function (Eq. \,\ref{eq:nbar_vs_t}), and the natural decay data is fitted to an exponential function.}
    \label{fig:cooling}
\end{figure}
\begin{figure*}[ht!]
    \centering
    \subfloat{\begin{overpic}[width=0.99\linewidth]{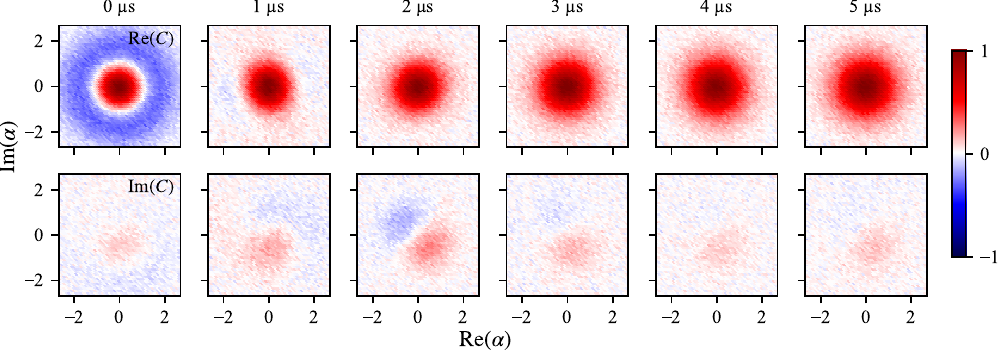}
    \put(-0.5,35){(a)}
    \end{overpic}}\\[5mm]
    \subfloat{\begin{overpic}[width=0.45\linewidth]{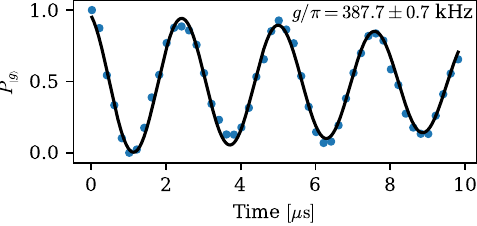}
    \put(0,50){(b)}
    \end{overpic}}
    \hspace{5mm}
    \subfloat{\begin{overpic}[width=0.45\linewidth]{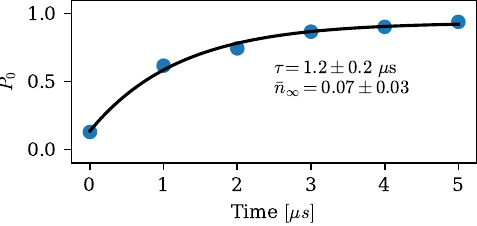}
    \put(0,50){(c)}
    \end{overpic}}
    
    \caption{Reset of a single Fock state by RDR. (a) The measured Wigner characteristic functions of the memory mode during the reset process. The initial single Fock state was generated with fidelity of $80\%$ and was cooled down to a vacuum state with fidelity of $93\%$. (b) Vacuum Rabi oscillations between the dressed qubit and the displaced memory mode. (c) Probability of the memory mode being in the vacuum state, extracted using maximum likelihood estimation of the Wigner characteristic functions. The probability is fitted to an exponential decay.}
    \label{fig:fock}
\end{figure*}

An application of RDR, as depicted in Fig.\,\ref{fig:cooling}(a), contains the following sequence: the sideband drives of both modes are ramped up, followed by the Rabi-drive. The drives are applied for a long enough time to fully reset the system. The Rabi-drive is ramped down and a $\pi/2$ pulse is applied to return the qubit to the bare qubit ground state. Finally, the sideband drives are ramped down. We used a ramp duration of $800\,\mathrm{ns}$.

To evaluate the performance of RDR, we carried out two reset experiments. In the first one, we cooled down an on-demand generated thermal state (see Appendix\,\ref{appendix:thermal}). For cooling, we set $\bar{a}_r=\kappa/\chi_r$ and sweep over $\bar{a}_m$ to find the optimal coupling rate between the qubit and the memory mode. To extract the average photon number after cooling, we measured the Wigner characteristic function of the memory mode along the real and imaginary axes. The average photon number was then obtained from the second derivative around the origin\,\cite{CharFuncNbar}
\begin{equation}\label{eq:nbar}
    \bar{n} = -\frac{1}{2}\left[\frac{\partial C(\alpha)}{\partial\alpha \partial\alpha^*}+\frac{\partial C(\alpha)}{\partial\alpha^* \partial\alpha}+1\right]_{\alpha=0}.
\end{equation}
To evaluate the derivative, we fitted the data around the origin to a $\mathrm{2^{nd}}$ order polynomial function. Naively, one could try to fit a Lorentzian function, as befits thermal states. However, non-Markovian effects cause the state to deviate from a thermal form during the cooling process. This is also reflected in the fact that $\bar{n}$ does not follow a purely exponential decay in Fig.\,\ref{fig:cooling}(c). Rather, the photon decay rate is bounded by $\mathcal{O}\left(\kappa\right)$, as only a single photon can populate the qubit at a time. This behavior is captured by a linear decay at high photon numbers and an exponential decay at low photon numbers. Therefore, the average photon number was fitted with a piecewise function of the form
\begin{equation}\label{eq:nbar_vs_t}
\bar{n}(t) =
\begin{cases}
A+B+B\gamma (t-t_0), &  t < t_0 \\
A+Be^{-\gamma(t-t_0)}, &  t \ge t_0
\end{cases},
\end{equation}
where $t_0$ is the time when the effective coupling rate is low enough such that the qubit no longer acts as a "bottle neck". Increasing the effective coupling beyond $\bar{a}_m\chi_m=\kappa/2$ only decreased the cooling rate (See Appendix\,\ref{appendix:beyondWeakCoupling}). The fitted maximum decay rates are $\frac{d\bar{n}}{dt}=-0.73 \pm 0.08\,\mathrm{MHz}$ for $\bar{a}_m\chi_m = \kappa/2$ and $\frac{d\bar{n}}{dt}=-0.44 \pm 0.05\,\mathrm{MHz}$ for $\bar{a}_m\chi_m = \kappa/8$. The steady state average photon number is $n_\infty = 0.045 \pm 0.025$ for the optimal cooling rate of $\bar{a}_m\chi_m = \kappa/2$. 

In the second experiment, we prepared a single Fock state and cooled it down using RDR. We set $\bar{a}_r=\kappa/2\chi_r$ to avoid non-Markovian effects. The Fock state was generated by applying the Rabi drive together with the sideband drive on the memory mode, which creates a JC interaction similar to the RDR protocol but without the sideband drive on the readout mode. We observed vacuum Rabi oscillations between the dressed qubit and the displaced memory mode (Fig.\,\ref{fig:fock}(b)). By applying the drives for a duration of $\pi/2\chi_m\bar{a}_m$ after initializing the qubit in the excited dressed-state, we successfully created a single Fock state in the memory mode. We also note that this method may be used to generate higher Fock states in a relatively fast manner.

\section*{Discussion}

The cooling rate of the RDR method is hard limited by the linewidth of the readout mode, as a further increase of the coupling rate would give rise to non-Markovian effects. Ideally, upon complete hybridization of the modes, the maximum cooling rate should be $\kappa/2$. Assuming a quasi-static steady state for the qubit and readout mode\,\cite{NatanTheory}, we can approximate the cooling rate to be limited to $\kappa/4$. In practice, we found that the maximum cooling rate was approximately $\kappa/3.3$ for $\bar{a}_m\chi_m = \kappa/2$ and $\kappa/5.45$ for $\bar{a}_m\chi_m = \kappa/8$.

Other less severe limitations arise from spurious transitions to parasitic two-level systems or to levels beyond the qubit subspace through higher-order interactions, that may be mediated by the strong drives\,\cite{spuriousTransitions}. These transitions may be prevented as the method allows for tuneabiliy of the Rabi frequency.
The Rabi frequency is tuneable in a range with an upper bound given by the available sideband drive power and the anharmonicity of the transmon and a lower bound given by $\chi_r,~\chi_m,~\bar{a}_r\chi_r,~\bar{a}_m\chi_m$ and $\kappa_r$, for the approximations to hold. The available power limits the Rabi frequency because $\bar{a}_m$ increases with $\epsilon_m$ and decreases with $\Omega_R$, according to Eq.\,\ref{eq:abar}. Therefore, setting $\Omega_R$ beyond the available range of $\epsilon_m$ would reduce the coupling and the cooling rate.

\section*{Conclusions}
We have demonstrated the Rabi-Driven Reset method, which enables rapid cooling of an electromag-
netic mode in a superconducting cavity via its coupling
to a transmon qubit and a readout resonator. The reset
time achieved with this protocol is more than two or-
ders of magnitude shorter than the natural decay time
of the system, and is only limited by the linewidth of
the readout mode. Importantly, the method provides
efficient cooling even in weakly coupled systems, where
higher-order effects such as self-Kerr nonlinearity remain
negligible. In such systems, our method is faster than all others to the best of our knowledge.

\section*{Funding}
This research was supported by the Israeli Science Foundation grant No. 657/23, and Technion’s Helen Diller Quantum Center.
\section*{Data availability}
All data supporting the findings of this study are available upon reasonable request.

\section*{Appendices}
\appendix
\section{System schematics}\label{appendix:system}
\begin{figure*}
    \centering
    \includegraphics[width=1\linewidth]{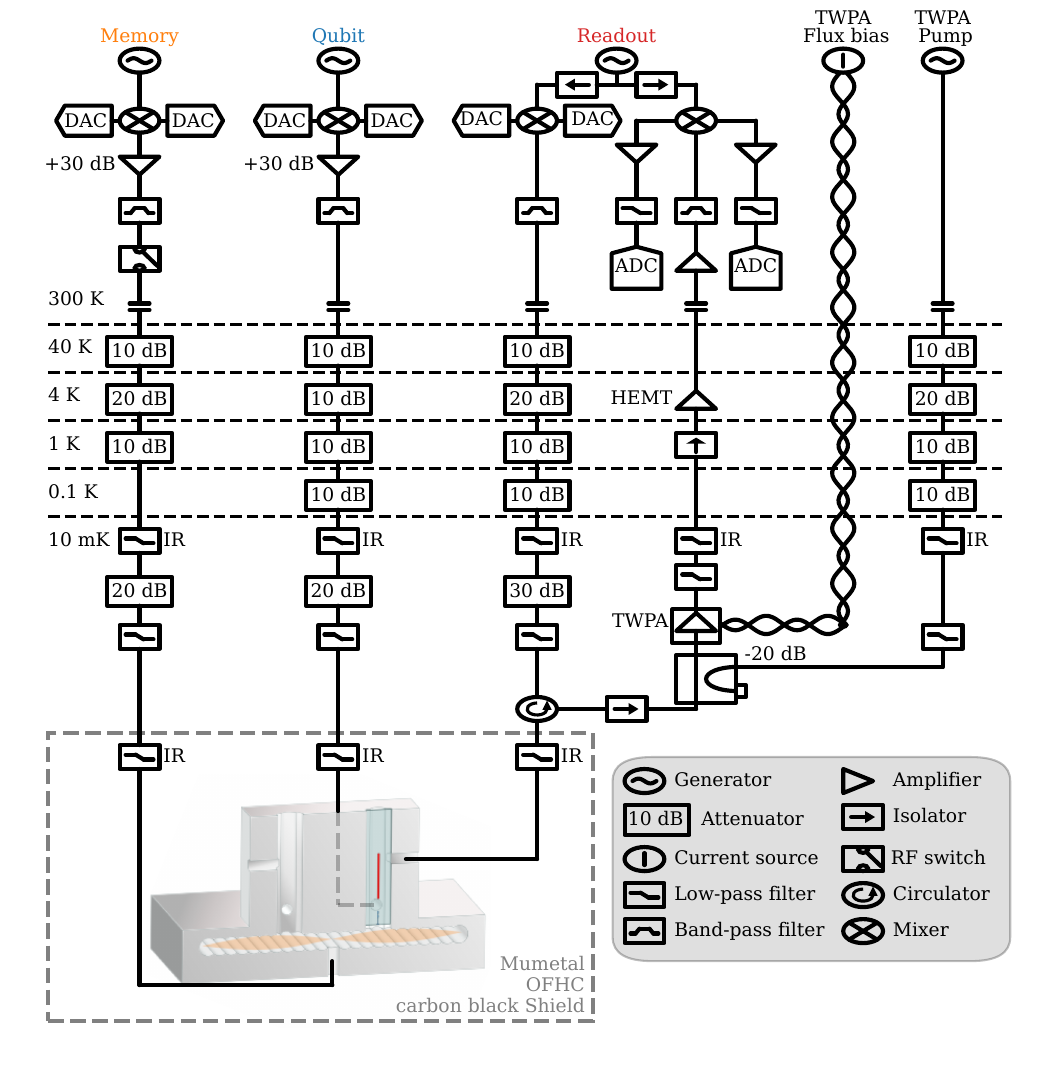}
    \caption{Full experimental system schematics.}
    \label{fig:system}
\end{figure*}

The full wiring diagram is presented in Fig.\,\ref{fig:system}.
\section{Sideband calibration}\label{appendix:calibration}

\begin{figure}
    \centering
    \includegraphics[width=\linewidth]{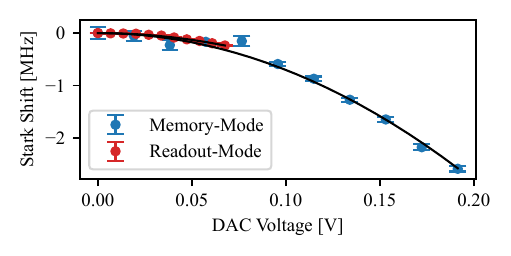}
    \caption{Stark-shift measurements as a function of drive.}
    \label{fig:driven_ramsey_calib}
\end{figure}

We calibrated the sideband amplitude by measuring the AC-Stark-shift induced by the sideband drive\,\cite{StarkShift} in a Ramsey experiment, as presented in Fig.\,\ref{fig:driven_ramsey_calib}. We sampled the detuned frequency as function of drive amplitude and extracted $\varepsilon$ from
\begin{equation}
    \delta\omega
    = \mathrm{Re}\left(\frac{8 \chi  \varepsilon ^2}{4 \Omega _R^2+(\kappa -2 i \chi )^2}\right)\approx 
    2\chi\left(\frac{\varepsilon}{\Omega_R}\right)^2.
\end{equation}
\section{Rabi calibration}\label{appendix:Rabi}
We needed to calibrate the Rabi frequency in a Stark-shifted frame. Naively, one would apply the sideband-drive to induce the Stark-shift and apply the Rabi-drive to measure its frequency. However, the generated JC interaction then highly complicates the calibration. Therefore, we have found an alternative way to conduct the calibration. If the Stark-shift $\delta$ is small compared to the target Rabi frequency $\Omega_R$ ($\Omega_R/2\pi=9\,\mathrm{MHz}$ and $\delta/2\pi=2.6\,\mathrm{MHz}$ with $\bar{a}_m\chi_m=\kappa/2$), one may simply apply the Rabi drive in a Stark-shifted frame and find the Rabi frequency from $\Omega=\sqrt{\Omega^2_R+\delta^2}$. But, if the induced Stark-shift is too large ($\Omega_R/2\pi=9\,\mathrm{MHz}$ and $\delta/2\pi=10.5\,\mathrm{MHz}$ with $\bar{a}_m\chi_m=\kappa$), the amplitude of the Rabi oscillations decreases, making the calibration noisy. To overcome this issue, one may use a sideband of a completely different frequency that induces the same Stark-shift. We used detuning of $5$ MHz with $\Omega_R/2\pi=9\,\mathrm{MHz}$. This way, the JC interaction is highly off-resonant and does not interfere with the observed Rabi oscillations. It is preferable to apply the sideband on the readout mode, where higher power is available and the dispersive coupling is larger.

\section{Beyond weak coupling}\label{appendix:beyondWeakCoupling}
\begin{figure}
    \centering
    \includegraphics[width=1\linewidth]{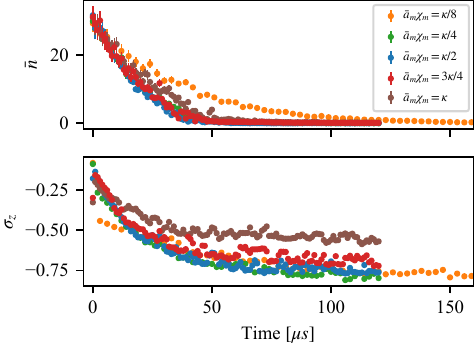}
    \caption{Cooling of thermal states as function of time. The top panel shows the average photon number and the bottom panel shows the expectation value along the Z axis of the effective qubit. This expectation value was extracted from the $M1$ measurement (see Fig.\,\ref{fig:cooling}(a).}
    \label{fig:MoreCooling}
\end{figure}

When cooling down a thermal state, we used a variety of memory-mode sideband amplitudes $\bar{a}_m$ to generate coupling rates between $0$ and $\kappa$. Fig.\,\ref{fig:cooling} (c) shows only three sideband-drive amplitudes, which coupling rates ranging up to $\kappa/2$, to avoid overcrowding of data points. Fig.\,\ref{fig:MoreCooling} shows the additional data, which clearly indicates that increasing the coupling beyond $\kappa/2$ did not increase the cooling rate and only decreased the purity of the steady state of the qubit.
\section{On-demand thermal state generation}\label{appendix:thermal}
We generated a thermal state by applying random displacements to the vacuum state. We sampled 1500 displacement amplitudes $\alpha$ from the distribution\,\cite{thermalDistribution}
\begin{equation}
    P(\left|\alpha\right|)=\frac{1}{\pi \bar{n}}e^{-\frac{\left|\alpha\right|^2}{\bar{n}}},
\end{equation}
where $\bar{n}$ is the average thermal photon number. We then loaded the amplitudes into the FPGA so it can sample a random amplitude by index in each iteration. The displacement phase was sampled evenly.

\bibliographystyle{apsrev4-2}
\bibliography{bib}

@article{fluteCavity2021,
  title = {Seamless High-$Q$ Microwave Cavities for Multimode Circuit Quantum Electrodynamics},
  author = {Chakram, Srivatsan and Oriani, Andrew E. and Naik, Ravi K. and Dixit, Akash V. and He, Kevin and Agrawal, Ankur and Kwon, Hyeokshin and Schuster, David I.},
  journal = {Phys. Rev. Lett.},
  volume = {127},
  issue = {10},
  pages = {107701},
  numpages = {6},
  year = {2021},
  month = {Aug},
  publisher = {American Physical Society},
  doi = {10.1103/PhysRevLett.127.107701},
  url = {https://link.aps.org/doi/10.1103/PhysRevLett.127.107701}
}

@article{highQCavity,
  title = {Superconducting Cavity Qubit with Tens of Milliseconds Single-Photon Coherence Time},
  author = {Milul, Ofir and Guttel, Barkay and Goldblatt, Uri and Hazanov, Sergey and Joshi, Lalit M. and Chausovsky, Daniel and Kahn, Nitzan and \ifmmode \mbox{\c{C}}\else \c{C}\fi{}ifty\"urek, Engin and Lafont, Fabien and Rosenblum, Serge},
  journal = {PRX Quantum},
  volume = {4},
  issue = {3},
  pages = {030336},
  numpages = {16},
  year = {2023},
  month = {Sep},
  publisher = {American Physical Society},
  doi = {10.1103/PRXQuantum.4.030336},
  url = {https://link.aps.org/doi/10.1103/PRXQuantum.4.030336}
}

@article{NatanTheory,
  title = {Cavity-mode initialization via a Rabi-driven qubit},
  author = {Karaev, N. and Blumenthal, E. and Moshel, G. and Diringer, A.A. and Hacohen-Gourgy, S.},
  journal = {Phys. Rev. Appl.},
  volume = {23},
  issue = {1},
  pages = {014043},
  numpages = {12},
  year = {2025},
  month = {Jan},
  publisher = {American Physical Society},
  doi = {10.1103/PhysRevApplied.23.014043},
  url = {https://link.aps.org/doi/10.1103/PhysRevApplied.23.014043}
}

@Article{ECDCooling,
author={Sivak, V. V.
and Eickbusch, A.
and Royer, B.
and Singh, S.
and Tsioutsios, I.
and Ganjam, S.
and Miano, A.
and Brock, B. L.
and Ding, A. Z.
and Frunzio, L.
and Girvin, S. M.
and Schoelkopf, R. J.
and Devoret, M. H.},
title={Real-time quantum error correction beyond break-even},
journal={Nature},
year={2023},
month={Apr},
day={01},
volume={616},
number={7955},
pages={50-55},
issn={1476-4687},
doi={10.1038/s41586-023-05782-6},
url={https://doi.org/10.1038/s41586-023-05782-6}
}

@Article{BadCooling,
author={Pfaff, Wolfgang
and Axline, Christopher J.
and Burkhart, Luke D.
and Vool, Uri
and Reinhold, Philip
and Frunzio, Luigi
and Jiang, Liang
and Devoret, Michel H.
and Schoelkopf, Robert J.},
title={Controlled release of multiphoton quantum states from a microwave cavity memory},
journal={Nature Physics},
year={2017},
month={Sep},
day={01},
volume={13},
number={9},
pages={882-887},
issn={1745-2481},
doi={10.1038/nphys4143},
url={https://doi.org/10.1038/nphys4143}
}

@Article{ECD,
author={Eickbusch, Alec
and Sivak, Volodymyr
and Ding, Andy Z.
and Elder, Salvatore S.
and Jha, Shantanu R.
and Venkatraman, Jayameenakshi
and Royer, Baptiste
and Girvin, S. M.
and Schoelkopf, Robert J.
and Devoret, Michel H.},
title={Fast universal control of an oscillator with weak dispersive coupling to a qubit},
journal={Nature Physics},
year={2022},
month={Dec},
day={01},
volume={18},
number={12},
pages={1464-1469},
issn={1745-2481},
doi={10.1038/s41567-022-01776-9},
url={https://doi.org/10.1038/s41567-022-01776-9}
}

@article{StarkShift,
  title = {Qubit-photon interactions in a cavity: Measurement-induced dephasing and number splitting},
  author = {Gambetta, Jay and Blais, Alexandre and Schuster, D. I. and Wallraff, A. and Frunzio, L. and Majer, J. and Devoret, M. H. and Girvin, S. M. and Schoelkopf, R. J.},
  journal = {Phys. Rev. A},
  volume = {74},
  issue = {4},
  pages = {042318},
  numpages = {14},
  year = {2006},
  month = {Oct},
  publisher = {American Physical Society},
  doi = {10.1103/PhysRevA.74.042318},
  url = {https://link.aps.org/doi/10.1103/PhysRevA.74.042318}
}

@article{MurchBath,
  title = {Cavity-Assisted Quantum Bath Engineering},
  author = {Murch, K. W. and Vool, U. and Zhou, D. and Weber, S. J. and Girvin, S. M. and Siddiqi, I.},
  journal = {Phys. Rev. Lett.},
  volume = {109},
  issue = {18},
  pages = {183602},
  numpages = {5},
  year = {2012},
  month = {Oct},
  publisher = {American Physical Society},
  doi = {10.1103/PhysRevLett.109.183602},
  url = {https://link.aps.org/doi/10.1103/PhysRevLett.109.183602}
}

@book{CharFuncNbar,
  title        = {Quantum Optics},
  author       = {Walls, D. F. and Milburn, G. J.},
  year         = {1994},
  edition      = {2nd},
  publisher    = {Springer},
  address      = {Berlin, Heidelberg},
  doi          = {10.1007/978-3-642-79548-0}
}

@article{DissipativeCat,
  title        = {Enhancing dissipative cat qubit protection by squeezing},
  author       = {Rémi Rousseau and Diego Ruiz and Emanuele Albertinale and Pol d'Avezac and Danielius Banys and Ugo Blandin and Nicolas Bourdaud and Giulio Campanaro and Gil Cardoso and Nathanael Cottet and Charlotte Cullip and Samuel Deléglise and Louise Devanz and Adam Devulder and Antoine Essig and Pierre Février and Adrien Gicquel and Élie Gouzien and Antoine Gras and Jérémie Guillaud and Efe Gümüş and Mattis Hallén and Anissa Jacob and Paul Magnard and Antoine Marquet and Salim Miklass and Théau Peronnin and Stéphane Polis and Felix Rautschke and Ulysse Réglade and Julien Roul and Jeremy Stevens and Jeanne Solard and Alexandre Thomas and Jean-Loup Ville and Pierre Wan-Fat and Raphaël Lescanne and Zaki Leghtas and Joachim Cohen and Sébastien Jezouin and Anil Murani},
  journal      = {arXiv preprint arXiv:2502.07892},
  year         = {2025},
  month        = feb,
  doi          = {10.48550/arXiv.2502.07892},
  url          = {https://arxiv.org/abs/2502.07892},
  eprint       = {2502.07892},
  archivePrefix= {arXiv},
  primaryClass = {quant-ph}
}

@article{spuriousTransitions,
  title        = {Spectroscopy of drive‐induced unwanted state transitions in superconducting circuits},
  author       = {W. Dai and S. Hazra and D. K. Weiss and P. D. Kurilovich and T. Connolly and H. K. Babla and S. Singh and V. R. Joshi and A. Z. Ding and P. D. Parakh and J. Venkatraman and X. Xiao and L. Frunzio and M. H. Devoret},
  journal      = {arXiv preprint arXiv:2506.24070},
  year         = {2025},
  doi          = {10.48550/arXiv.2506.24070}
}

@article{thermalDistribution,
  title = {Thermal coherent states in the Bargmann representation},
  author = {Vourdas, A. and Bishop, R. F.},
  journal = {Phys. Rev. A},
  volume = {50},
  issue = {4},
  pages = {3331--3339},
  numpages = {0},
  year = {1994},
  month = {Oct},
  publisher = {American Physical Society},
  doi = {10.1103/PhysRevA.50.3331},
  url = {https://link.aps.org/doi/10.1103/PhysRevA.50.3331}
}

@article{drachma,
  title        = {Dispersive Qubit Readout with Intrinsic Resonator Reset},
  author       = {Jerger, M. and Motzoi, F. and Gao, Y. and Dickel, C. and Buchmann, L. and Bengtsson, A. and Tancredi, G. and Warren, C. W. and Bylander, J. and DiVincenzo, D. and Barends, R. and Bushev, P. A.},
  journal      = {arXiv preprint arXiv:2406.04891},
  year         = {2024},
  eprint       = {2406.04891},
  archivePrefix = {arXiv},
  primaryClass = {quant-ph},
  doi          = {10.48550/arXiv.2406.04891}
}

\section*{Contributions}
The project was conceived by E.B. The transmon qubit was fabricated by E.B. and N.K. The measurements and data analysis were conducted by E.B. and N.K. The project was supervised by S.H.G. The manuscript was written by E.B. All authors discussed the results and the manuscript.

\section*{Ethics declarations}

The authors declare no competing interests.

\end{document}